\documentclass{ws-p9-75x6-50}

\begin{document}

\title{Observational Approaches to the Topology of the Universe}

\author{B. F. Roukema}

\address{Inter-University Centre for Astronomy and Astrophysics, 
 Post Bag 4, Ganeshkhind, Pune, 411 007, India\\
 Email: boud@iucaa.ernet.in}


\maketitle

\abstracts{
Many different and complementary strategies
for translating the basic principle of multiple topological imaging into
observational analysis are now available,
both for three-dimensional and two-dimensional catalogues.}

\section{Introduction}
Observational cosmic topology\cite{LaLu95,LR99}
shares the fundamental observational
problems of curvature estimates (evolution of objects, 
peculiar velocities, etc., see Section 5.3 of ref.\cite{LR99}), 
but since the former concerns global
geometry and the latter only local geometry, the former requires additional 
work in classifying the methods from a purely geometrical point
of view. Most work is based on the principle of 
the existence of multiple topological images of single physical objects.


\section{Complementary geometrical strategies}

{
\begin{list}{\Alph{enumi}.}{\usecounter{enumi}} \parskip=-8pt
\item multiple topological images:
  \begin{list}{\Alph{enumi}.\roman{enumii}}{\usecounter{enumii}}
     \item three-dimensional (collapsed astrophysical objects):
  \begin{list}
{\Alph{enumi}.\roman{enumii}.\arabic{enumiii}}{\usecounter{enumiii}}
     \item local isometries\cite{Rouk96,ULL99,LUL00} - multiple occurrence 
of ``type I pairs'' or ``local pairs''
     \item cosmic 
crystallography\cite{LLL96,LLU98,G99a,FagG97,FagG99a,G99b,G99c,ULL99,LUL00}
 - multiple occurrence of ``type II pairs'' or ``generator pairs'',
i.e. of pairs of objects in comoving space separated by a generator
     \item[$\bullet$] for comparison, the uncorrelated pairs
from simply connected random 
simulations could be called ``type III pairs'' or ``random pairs''
     \item characteristics of individual objects\cite{Gott80,RE97,Wich99}
     \end{list}

     \item two-dimensional (microwave background, CMB):
     \begin{list}
{\Alph{enumi}.\roman{enumii}.\arabic{enumiii}}{\usecounter{enumiii}}
	\item identified circles principle: discovery of 
principle\cite{Corn96,Corn98b} and its quantitative application to COBE 
data\cite{Rouk00a,Rouk00c}
	\item patterns of spots\cite{LevSGSB98}
	\item perturbation statistics assumptions: see refs listed under 
(i) in Section 1.2 of ref.\cite{Rouk00a}; 
controversy 
against\cite{BPS98} and
in favour of\cite{Inoue99,Aur99,CS99} 
hyperbolic compact 3-manifolds 
by applying these assumptions to COBE data remains
     \end{list}
  \end{list}
\item other:
  \begin{list}{\Alph{enumi}.\roman{enumii}}{\usecounter{enumii}}
    \item cosmic strings\cite{UzPet97}
    \item nested crystallography\cite{Rouk00b}
  \end{list}
\end{list}
}  



\def\apj{ApJ.}                        
\def\apjs{ApJ. Supp.}                 
\def\aj{A.J.}                           
\def\aanda{A\&A}                      
\def\cqg{Class. Quant. Grav.}         
\def\mnras{Month. Not. R. Astron. Soc.}

\newcommand\joref[5]{#1, #5, {#2, }{#3, } #4}
\newcommand\epref[3]{#1, #3, {#2}}


\end{document}